\documentclass[a4paper, 11pt]{article}

\usepackage{xcolor}
\usepackage{setspace}
\usepackage{hanging}
\usepackage{changepage}
\usepackage{hyperref}
\definecolor{mycolor}{RGB}{0,88,204}
\hypersetup{
  colorlinks=true,
  linkcolor=black,
  urlcolor=mycolor,
  citecolor=mycolor
}
\usepackage{amsmath}
\usepackage{amsthm}
\usepackage{tikz}
\theoremstyle{definition}

\newcounter{mycounter}[section]









\usepackage{geometry}

\usepackage{dirtree}
\usepackage{fancyhdr}

\fancypagestyle{plain}{%
  \fancyhf{}
  \fancyhead[R]{\thepage}
  
}

\usepackage{mathptmx}
\DeclareMathAlphabet{\mathcal}{OMS}{cmsy}{m}{n}

\usepackage{titlesec}

\titleformat{\section}
  {\normalfont\large\centering}{\thesection.}{1em}{\MakeUppercase}
\titleformat{\subsection}
  {\normalfont\centering}{\thesubsection.}{1em}{\MakeUppercase}

  \titlespacing{\paragraph}{10pt}{0pt}{6pt}[0pt]

\usepackage{listings}
\usepackage[T1]{fontenc}
\usepackage[utf8]{inputenc}
\usepackage{amssymb}
\usepackage{chngcntr}
\usepackage{inconsolata}
\usepackage{upquote}
\definecolor{keywordcolor}{rgb}{0.7, 0.1, 0.1}   
\definecolor{tacticcolor}{rgb}{0.0, 0.1, 0.6}    
\definecolor{commentcolor}{rgb}{0.4, 0.4, 0.4}   
\definecolor{symbolcolor}{rgb}{0.0, 0.1, 0.6}    
\definecolor{sortcolor}{rgb}{0.1, 0.5, 0.1}      
\definecolor{attributecolor}{rgb}{0.7, 0.1, 0.1} 

\lstnewenvironment{code}[1][]%
{
   \noindent\newline
   \minipage{1\linewidth}
   \vspace{0.5\baselineskip}
   \lstset{
 	frame = lrtb,
 	rulecolor=\color{mycolor},
 	escapeinside={/*!}{!*/},
	language=lean,
	aboveskip = 5mm,
	belowskip = 5mm,
	xleftmargin=2mm,
	xrightmargin=2mm,
	}
	}
{\endminipage\newline}
\lstnewenvironment{codeLong}[1][]%
{
 \lstset{
	frame = lrtb,
	rulecolor=\color{mycolor},
	escapeinside={/*!}{!*/},
	language=lean,
	aboveskip = 5mm,
	belowskip = 5mm,
	xleftmargin=2mm,
	xrightmargin=2mm,
	numbers=left,
	}
	}
{}

\usepackage{xpatch}
\usepackage{listings}
\usepackage{realboxes}
\definecolor{mycolorSubtle}{RGB}{245,250,255}
\DeclareRobustCommand{\myinline}{\lstinline}
\makeatletter
\xpretocmd\myinline{\Colorbox{mycolorSubtle}\bgroup\appto\lst@DeInit{\egroup}}{}{}
\makeatother
\newcommand{\codeLink}[1]{
  \vspace{-0.5cm}\hfill\href{https://github.com/HEPLean/HepLean/blob/1b951994ae3d882242b02d23957ef1ee7fa05f3d/HepLean/#1}{(source)}
  }
  
 \newcommand{\textLink}[1]{\href{https://github.com/HEPLean/HepLean/blob/1b951994ae3d882242b02d23957ef1ee7fa05f3d/HepLean/#1}{source}}
 \newcommand{\textLinkB}[1]{\href{https://github.com/HEPLean/HepLean/blob/1b951994ae3d882242b02d23957ef1ee7fa05f3d/HepLean/#1}{(source)}}
\title{Digitalizing Wick's theorem}
\author{Joseph Tooby-Smith \\ \textit{Reykjavik University, Reykjavik, Iceland}}
\date{}

\begin{document}
\counterwithin{lstlisting}{section}
\maketitle
\vspace{-1cm}
\begin{abstract}
	Wick's theorem is a cornerstone of perturbative quantum field theory.
	In this paper we announce and discuss the digitalization of Wick's theorem 
	and its proof into the interactive theorem prover Lean~4 as part 
	of the project PhysLean. 
	We do the same for the static and normal-ordered versions of Wick's theorem.
\end{abstract}
\tableofcontents

\section{Introduction}

The operator algebra associated with a quantum field theory 
has two linear maps associated with it, called normal-ordering and time-ordering.
These two maps rearrange products of field operators. 
Wick's theorem comes in three versions, which are statements about the relationship between these two maps:
the static version relates an ordinary product of field operators 
to a sum of normal-ordered products; 
the standard version relates a time-ordered product of field operators 
to a sum of normal-ordered products; 
and the normal-ordered version relates a product of field operators that is first normal-ordered then time-ordered
to a sum of normal-ordered products.

In this setting Wick's theorem~\cite{wick} 
(see also e.g.~\cite{molinari2023noteswickstheoremmanybody}) is the 
precursor to Feynman diagrams, which form a bedrock
to perturbative quantum field theory.
In mathematics, and in particular probability theory, it is related to 
Isserlis' theorem, which allows 
the computation of higher-order moments in a multivariate normal distribution. 

We have formalized (or, to prevent confusion with the other 
meaning of `formalized' prominant in physics, digitalized)
the three versions of Wick's theorem and their proofs into the interactive theorem prover Lean 4~\cite{lean}.
As far as we are aware, this is the first time that the Wick's theorems have been
formalized into any theorem prover.

For the reader new to theorem provers: They are computer programs 
that check statements and proofs of results for mathematical correctness, using 
some mathematical foundation, for example type theory.

Our choice of the specific theorem prover Lean is due to 
the extensive, and growing, library of formalized mathematics 
called Mathlib~\cite{mathlib}.
Our formalization of Wick's theorem builds upon this library,
but is integrated, due to its prominant application in physics and its framing, 
into the library Physlean~\cite{PhysLean} (formerly called HepLean~\cite{HepLean}).
The mission of PhysLean is to create an open-source, community built, library of formalized results from 
physics in Lean 4, with a  focus on useability to the wider physics community.
Other projects related to science and Lean include work on 
chemical physics~\cite{josephson} and work on scientific computing in Lean~\cite{SciLean}.

Despite living outside of Mathlib, PhysLean and this formalization of the Wick's theorems follows 
the conventions of Mathlib. In particular we have formalized many 
extra lemmas and definitions around Wick's theorem (sometimes called an API)
to assist in future applications.
The result of this is that the part of PhysLean 
dedicated to Wick's theorem and related results,
spans 52 files, has 193 definitions and 929 lemmas. The relevant directory structure is as follows:
\\
\noindent\rule{\textwidth}{0.4pt}
\dirtree{%
.1 PerturbationTheory.
.2 FieldOpFreeAlgebra.
.3 $\ldots$.
.2 FieldSpecification.
.3 $\ldots$.
.2 FieldStatistics.
.3 $\ldots$.
.2 Koszul.
.3 $\ldots$.
.2 WickAlgebra.
.3 $\ldots$.
.2 WickContraction.
.3 $\ldots$.
.2 CreateAnnihilate.lean.
}
\noindent\rule{\textwidth}{0.4pt}

To answer the question of what the point of this formalization is, 
we note that formalizing physics into the open-source community-run project PhysLean has a number of benefits. 
The first, and most obvious, is the confidence it provides in the correctness of the results, 
and their proofs. But this is only one of the many benefits.
Secondly, it allows results to be reused to prove further results,
and in this sense, everything need only be formalized once. 
Likewise, improvements made through refactoring can be propagated through the whole 
project. 
Thirdly, due to the program knowing the underlying mathematical structure of the objects 
involved, the library becomes highly structured and searchable (see e.g.~loogle~\cite{loogle}).
Fourthly, there is ongoing research (see e.g.~\cite{google, LeanDojo}) into the use of 
machine learning to assist theorem proving in Lean. 
As these machine learning tools improve, it will be possible to use them in PhysLean to 
accelerate formalization and potentially the discovery of results. 
Fifthly, many parts of the library are computable functions and therefore can 
directly be used in a computer program, allowing a new interface between 
theorem proving and programming.
Lastly, the whole project is a new type of open-source collaboration
within physics, and has the possibility to include contributions from undergraduate and master student projects, 
aiding in pedagogy. This can be combined with structured notes and games created 
from the project.

From the physics side, it is hoped that our formalization of Wick's theorem
can serve as a foundation to and be widely 
applied in the formalization of perturbative quantum field theory.
In particular, we hope it will be useful in the formalization of Feynman diagrams, 
which will be the subject of a future paper.

In this paper we will not give the full details of the formalization, 
we will assume that the interested reader will look at and directly interact 
with the code, which has been well-documented with this in mind. 
The code can be found as part of the \myinline|PhysLean.QFT.PerturbationTheory| directory of:\footnote{Instructions for installing PhysLean can be found at: \url{https://physlean.com/GettingStarted}}
\begin{center}
	\url{https://github.com/HEPLean/PhysLean/releases/tag/v4.18.0}
\end{center}
Instead, we will give a high-level overview of the three forms of Wick's theorem 
formalized into PhysLean. The aim is to convey the broad
ideas behind the formalism. This encapsulates the main mathematical insight of this paper: 
a formulation of Wick's theorem adaptable for theorem provers. 

To orientate the computer science reader, the main data structure 
used in the formalization are Wick contractions, which are discussed in 
Section~\ref{sec:wickcontractions}.
\section{The Algebra}
\begin{figure}
\centering
\begin{tikzpicture}
	\node[align=center, name=stat] at (0, 1) {\myinline|FieldStatistic|};
	\node[align=center, name=createAnnih] at (8, 1) {\myinline|CreateAnnihilate|};
	\node[align=center, name=spaceTime] at (4, 2) {\myinline|Momentum ⊕ SpaceTime ⊕ Momentum|};
	
	\node[align=center, name=field] at (0,0) {\myinline|Field|};
	\node[align=center, name=fieldop] at (4,0) {\myinline|FieldOp|};
	\node[align=center, name=cranfieldop] at (8,0) {\myinline|CrAnFieldOp|};
	
	\node[align=center, name=o] at (0,-2) {\myinline|𝓞|};
	\node[align=center, name=WickAlgebra] at (4,-2) {\myinline|WickAlgebra|};
	\node[align=center, name=fieldopfreealgebra] at (8,-2) {\myinline|FieldOpFreeAlgebra|};

	\draw[->] (field) -- (stat);
	\draw[->] (fieldop) -- (spaceTime);
	\draw[->] (cranfieldop) -- (createAnnih);

	\draw[<-] (field) -- (fieldop);
	\draw[<-] (fieldop) -- (cranfieldop);
	
	\draw[->, dashed] (WickAlgebra) -- (o);
	\draw[->] (fieldopfreealgebra) -- (WickAlgebra);

	\draw[->] (cranfieldop) -- (fieldopfreealgebra);
	\draw[->] (fieldop) -- (fieldopfreealgebra);
\end{tikzpicture}
\caption{The types and algebras involved in the formalization of Wick's theorem.}
\label{fig:types}
\end{figure}
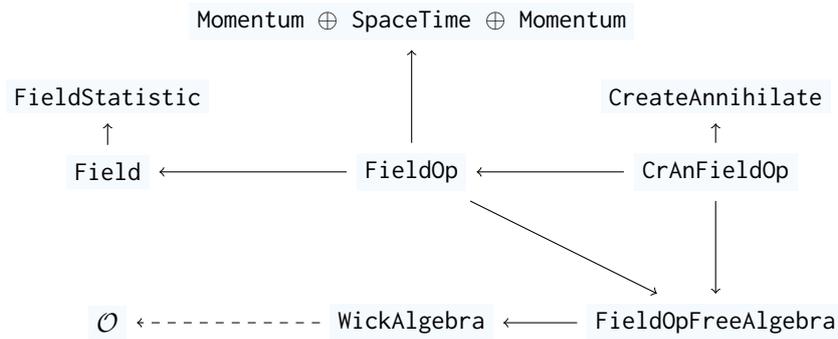

As stated in the introduction, Wick's theorem is usually formulated within 
the operator algebra \myinline|𝓞| acting on a Hilbert space associated with a quantum field theory.
We will not actually prove Wick's theorem in this algebra. 
Instead, we will work in a more general algebra called \myinline|WickAlgebra|. 
The algebra  \myinline|WickAlgebra| is the algebra with the minimal 
assumptions needed to prove Wick's theorem, and satisfies 
the relevant universality conditions to ensure that Wick's theorem in \myinline|𝓞| (however it is defined) is a simple
consequence thereof. 
In this section we will give the construction of \myinline|WickAlgebra|. 
The types and algebras involved are summarized in Figure~\ref{fig:types}.

Let us first define a number of fundmental types. 
A type can be thought of as roughly similar to a set.
The type \myinline|FieldStatistic| is an inductive type containg two elements 
\myinline|bosonic| and \myinline|fermionic|. This will govern the presence of 
certain signs in Wick's theorem and the need for commutators vs anti-commutators.
It and its properties are defined in the \myinline|<FieldStatistics>| directory. 

The type \myinline|CreateAnnihilate| is also an inductive type containing two elements
\myinline|create| and \myinline|annihilate|. It is defined 
in the \myinline|CreateAnnihilate.lean| file.
This type will govern the normal-ordering 
map which we will define in the next section. The type \myinline|Momentum| is
isomorphic to \myinline|ℝ³|, and the type \myinline|SpaceTime| is isomorphic to \myinline|ℝ⁴|, 
however from a mathematics point of view, the exact details of these two types is not important. 

The type \myinline|Field|, and to some extent \myinline|FieldOp|, 
contain the input to Wick's theorem. 
Mathematically, \myinline|Field| can be any type, however, from physics it should 
correspond to the type of fields present in the theory and is usually finite. We equip 
\myinline|Field| with a map to \myinline|FieldStatistic| which 
assigns a statistic to each field. The type \myinline|Field| is part of the \myinline|<FieldSpecification>| and 
is defined in the corresponding directory.

The type \myinline|FieldOp| physically corresponds to the field operators in the theory. 
This includes position-based operators, and incoming and outgoing asymptotic operators. 
It is determined by the fields present, what Lorentz representation they are in, 
whether they are real or complex etc. It is also defined in the \myinline|<FieldSpecification>| directory. 

Mathematically \myinline|FieldOp| could be taken as any type equipped with a map to \myinline|Field|, 
and a map to 
\myinline|Momentum ⊕ SpaceTime ⊕ Momentum|. 
Given two types \myinline|A| and \myinline|B|, the type \myinline|A ⊕ B| is 
the type whose elements, or terms, are either terms of \myinline|A| or terms of \myinline|B|.  
If a field operator \myinline|φ|
lands on the first \myinline|Momentum|, it is called an incoming asymptotic operator,
if it lands on the second \myinline|Momentum|, it is called an outgoing asymptotic operator,
and if it lands on the \myinline|SpaceTime|, it is called a position operator.
Our exact implementation of \myinline|FieldOp| sits between the physical and mathematical 
approaches, the details of which are not important for this discussion.

\sloppy We define, in the
\myinline|<FieldSpecification>| directory, the type  \myinline|CrAnFieldOp| to contain 
the creation and annihilation \emph{parts} of field operators.
It is defined with two maps, one to \myinline|FieldOp| and one to \myinline|CreateAnnihilate|, 
such that the following conditions hold.
The preimage in 
\myinline|CrAnFieldOp| of a position operator \myinline|φ| in \myinline|FieldOp|
contains exactly two elements, one of which maps to \myinline|create| in \myinline|CreateAnnihilate| 
(we say is a creation operator) and one of which maps to \myinline|annihilate| in \myinline|CreateAnnihilate|
(we say is an annihilation operator).
The preimage in
\myinline|CrAnFieldOp| of an incoming asymptotic operator \myinline|φ| in \myinline|FieldOp|
contains exactly one element, which maps to \myinline|create| in \myinline|CreateAnnihilate|.
Lastly, the preimage in
\myinline|CrAnFieldOp| of an outgoing asymptotic operator \myinline|φ| in \myinline|FieldOp|
contains exactly one element, which maps to \myinline|annihilate| in \myinline|CreateAnnihilate|.
This definition is a reflection of the creation and annihilation parts of field operators
in quantum field theory.  

We are now in a position to talk about algebras. The algebra \myinline|FieldOpFreeAlgebra|, 
despite its name, is the
free \myinline|ℂ|-algebra generated
by \myinline|CrAnFieldOp| (not by \myinline|FieldOp|).
It and its properties are defined in the \myinline|<FieldOpFreeAlgebra>| directory.
There is a map from \myinline|FieldOp| to \myinline|FieldOpFreeAlgebra|, 
defined such that \myinline|φ| is taken to the sum over elements in 
its preimage in \myinline|CrAnFieldOp|.
By virtue of its definition, \myinline|FieldOpFreeAlgebra| is spanned by lists of elements of \myinline|CrAnFieldOp|. 

On  \myinline|FieldOpFreeAlgebra| we have a bilinear map
\myinline|superCommutator|, which we define as follows.
We say a single element of \myinline|CrAnFieldOp| is \myinline|bosonic| or \myinline|fermionic|
if its underlying \myinline|Field| is. 
A list of elements of \myinline|CrAnFieldOp| is called \myinline|fermionic| 
if it has an odd number of \myinline|fermionic| elements, and \myinline|bosonic| otherwise.
The supercommutator is defined such that it is the anti-commutator on 
two lists which are \myinline|fermionic| and the commutator otherwise. 

For Wick's theorem to hold, we must impose certain conditions on the supercommutator. These are 
\begin{enumerate}
	\item the supercommutator of two elements of \myinline|CrAnFieldOp| which are both creation operators is zero;
	\item the supercommutator of two elements of \myinline|CrAnFieldOp| which are both annihilation operators is zero;
	\item the supercommutator of two elements of \myinline|CrAnFieldOp| one of which is bosonic and the other
		is fermionic is zero;
	\item the supercommutator of an element of \myinline|CrAnFieldOp| with the supercommutator of two other
		elements of \myinline|CrAnFieldOp| (corresponding the statement 
		that supercommutators are in the center) is zero.
\end{enumerate}
The supercommuators in these conditions generate an ideal of \myinline|FieldOpFreeAlgebra|. 
The algebra \myinline|WickAlgebra| is defined as the quotient of \myinline|FieldOpFreeAlgebra| by this ideal.
We will work with \myinline|WickAlgebra| in the rest of this paper. 
The directory \myinline|<WickAlgebra>| contains the definition of \myinline|WickAlgebra| and its properties.

The supercommuator turns out to be well-defined on the quotient \myinline|WickAlgebra|, 
and within this context we will denote it as \myinline|[..., ...]ₛ|.

\section{Time and normal ordering}

We now define the two linear maps on \myinline|WickAlgebra| called time-ordering 
and normal-ordering, which will play a vital role in Wick's theorem. 

Given any ordering relation on \myinline|CrAnFieldOp|, 
we can define a linear map from \myinline|FieldOpFreeAlgebra| to itself. 
This is done by taking a list of elements of \myinline|CrAnFieldOp| (which recall span \myinline|FieldOpFreeAlgebra|),
to the list ordered using the insertion-sort
algorithm, and multiplying them by a sign ($\pm 1$) corresponding to the number of
fermionic-fermionic exchanges needed to move the elements of the list into the correct order.
The properties of such a sign are defined as part of the \myinline|<Koszul>| directory of the project.

On \myinline|CrAnFieldOp| we have two natural orderings. The first is time-ordering, denoted \myinline|timeOrderRel|.
This is defined such that \myinline|timeOrderRel φ1 φ2| is true if
\myinline|φ1| has a \emph{greater} or equal time than \myinline|φ2|. 
More precisely, it is true if, 
\myinline|φ1| is an outgoing asymptotic operator (terminology inherited from \myinline|FieldOp|), 
\myinline|φ2| is an incoming asymptotic operator,
or \myinline|φ1| and \myinline|φ2| are both position operators and the time 
(the first element of \myinline|SpaceTime|) of \myinline|φ1| is greater than or equal to the time of \myinline|φ2|.
The other relation is normal-ordering, denoted \myinline|normalOrderRel|.
This is defined through \myinline|CreateAnnihilate| and such that \myinline|normalOrderRel φ1 φ2| for
\myinline|φ1| and \myinline|φ2|  in \myinline|CrAnFieldOp|
is true if \myinline|φ1| is a creation operator or \myinline|φ2| is an annihilation operator. 

\sloppy As stated above these two orderings define linear maps from \myinline|FieldOpFreeAlgebra| to itself.
These are given in the \myinline|<FieldOpFreeAlgebra>| directory.
The time-ordering and the normal-ordering linear maps on \myinline|WickAlgebra| are 
defined, in the \myinline|<WickAlgebra>| directory, by lifting these maps from \myinline|FieldOpFreeAlgebra| to \myinline|WickAlgebra|.
This can be done because they are well-defined with respect to the quotient, the proof 
of which is naturally formalized.
The time-ordering and normal-ordering maps on \myinline|WickAlgebra| 
are denoted \myinline|𝓣(a)| and \myinline|𝓝(a)| respectively, for \myinline|a| in \myinline|WickAlgebra|.

Given a list of elements \myinline|φs| of \myinline|FieldOp| (which one can think 
of, or `cast', as an element of \myinline|WickAlgebra|),
the static version of Wick's theorem is a statement about  \myinline|φs| directly, 
the standard version is a statement about \myinline|𝓣(φs)|, and 
the normal-ordered version is a statement about \myinline|𝓣(𝓝(φs))|.

In practice more general orderings could be used here. We have refrained from 
doing so to keep the formalization close to classical approaches and because 
Wick's theorem is almost exclusively used in the forms given. 

Before we give the versions of Wick's theorem, we need to define the notion of a Wick contraction.

\section{Wick contractions} \label{sec:wickcontractions}
To start with an example, for \myinline|[φ0, φ1, φ2, φ3, φ4, φ5]| a list of elements of \myinline|FieldOp|
(note not \myinline|CrAnFieldOp| here), 
a physicist would draw the following diagram to represent an example of a Wick contraction
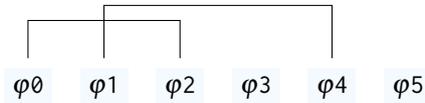
\begin{equation*}
\begin{tikzpicture}[baseline=(current bounding box.center)]
	\coordinate (0) at (0,0);
	\coordinate (1) at (1,0);
	\coordinate (2) at (2,0);
	\coordinate (3) at (3,0);
	\coordinate (4) at (4,0);
	\coordinate (5) at (5,0);

	\node[below] at (0,0) {\myinline|φ0|};
	\node[below] at (1,0) {\myinline|φ1|};
	\node[below] at (2,0) {\myinline|φ2|};
	\node[below] at (3,0) {\myinline|φ3|};
	\node[below] at (4,0) {\myinline|φ4|};
	\node[below] at (5,0) {\myinline|φ5|};

	\draw (0,0) -- (0,0.5) -- (2,0.5) -- (2,0);
	\draw (1,0) -- (1,0.7) -- (4,0.7) -- (4,0);
\end{tikzpicture}
\end{equation*}

Formally, in PhysLean, we define the type of Wick contractions, 
denoted \myinline|WickContraction φs.length|, associated 
with a list of elements \myinline|φs| of \myinline|FieldOp|, such 
that each element corresponds to a finite set of unordered disjoint pairs of positions 
in \myinline|φs|. For example, the Wick contraction above corresponds to
\myinline|{{0, 2}, {1, 4}}|. We typically denote Wick contractions 
associated with \myinline|φs| as \myinline|φsΛ|.
This definition and the properties of Wick contractions are in the 
the \myinline|<WickContraction>| directory of the project.

From a computer science point of view, the Wick contraction is the main data structure 
used in the formalization of Wick's theorem.

Note that the definition of Wick contractions only 
depends on the length of the list \myinline|φs|. 
This is done to allow Wick contractions to be computable, 
allowing users to calculate the Wick contractions of a list of \myinline|FieldOp|, 
and properties thereof,
with relative ease. This is part of a more general approach in PhysLean 
towards computability. 

Associated with Wick contractions are a number of properties which we now give in turn. 

The first property is the simplest. Given 
a list of elements \myinline|φs| of \myinline|FieldOp| with
a Wick contraction \myinline|φsΛ|, 
we get another list, denoted \myinline|[φsΛ]ᵘᶜ| of elements of \myinline|FieldOp|,
consisting of the elements of \myinline|φs| uncontracted in \myinline|φsΛ|.
For example, for the Wick contraction this list of uncontracted elements is \myinline|[φ3, φ5]|. 

The second property is the sign of the Wick contraction.
Informally, the sign ($\pm 1$) corresponds to the number of fermionic-fermionic exchanges
one needs to move the paired elements of the Wick contraction next to each other.
In PhysLean, we define the sign by moving paired elements next to each other 
starting from the leftmost Wick contraction. For the example above, 
assuming all fields are fermionic, we pick up a sign of $-1$ moving \myinline|φ2| through \myinline|φ1| 
to put \myinline|φ0| and \myinline|φ2| next to each other, and then a sign of $-1$ moving \myinline|φ4|  
through \myinline|φ3| to put \myinline|φ1| and \myinline|φ4| next to each other.
This gives a total sign of $+1$ for the Wick contraction.
If the Wick contraction only pairs bosons with bosons and fermions with fermions,
this choice of ordering does not matter, and in the formalization 
we have lemmas which correspond to this. 
Given a Wick contraction \myinline|φsΛ| of a list of elements \myinline|φs| of \myinline|FieldOp|,
the sign of the Wick contraction is denoted \myinline|φsΛ.sign|.

For a Wick contraction \myinline|φsΛ| of a list of elements \myinline|φs| of \myinline|FieldOp|,
the third property is an element of the center of \myinline|WickAlgebra| called \myinline|φsΛ.staticContract|.
This is defined as the product over contracted pairs of the supercommutor  
of the annihilation part of the first element of the pair with the whole of the second element of the pair.
Denoting the annihilation part in \myinline|WickAlgebra| of an element \myinline|φ| of \myinline|FieldOp| as \myinline|anPart φ|
(defined to be zero if \myinline|φ| is an incoming asymptotic operator),
in the above example \myinline|φsΛ.staticContract| corresponds to 
\myinline|[anPart φ0, φ2]ₛ * [anPart φ1, φ4]ₛ|. Both supercommuators are 
in the center of the algebra so order here does not matter. 

Related to the static contract is the notion of a time contract.
Given two elements \myinline|φ1| and \myinline|φ2| of \myinline|FieldOp|,
the time contract of \myinline|φ1| and \myinline|φ2|, denoted 
\myinline|WickAlgebra.timeContract φ1 φ2|, is defined as the element of \myinline|WickAlgebra|
given by 
\myinline|𝓣(φ * ψ) - 𝓝(φ * ψ)|. This time contraction is in the center of \myinline|WickAlgebra|. 
For a Wick contraction \myinline|φsΛ| of a list of elements \myinline|φs| of \myinline|FieldOp|,
the time contract of \myinline|φsΛ|, denoted \myinline|φsΛ.timeContract|,
is defined as the product over contracted pairs of the time contract of each pair.  
Thus, for the example above \myinline|φsΛ.timeContract| corresponds to
\myinline|WickAlgebra.timeContract φ0 φ2 * WickAlgebra.timeContract φ1 φ4|. 

As part of the API which makes up the formalization of Wick's theorem, 
we relate these properties of the Wick contractions to constructors thereof. 
One such constructor is \myinline|insertAndContract|. Given a list of field operators 
\myinline|φs| and a Wick contraction \myinline|φsΛ| of \myinline|φs|, 
another field operator \myinline|φ|, 
a position \myinline|i| in \myinline|φs|, and an optional position \myinline|k| in \myinline|[φsΛ]ᵘᶜ| 
(that is either \myinline|none| or \myinline|some k'| for some \myinline|k'| a position of \myinline|[φsΛ]ᵘᶜ|),
we define \myinline|φsΛ.insertAndContract φ i k|, denoted as \myinline|φsΛ ↩Λ φ i k|,
to be the Wick contraction of the list \myinline|φs| with \myinline|φ| inserted at position \myinline|i|,
and optionally contracted with the element at position \myinline|k| in \myinline|[φsΛ]ᵘᶜ|.
Then, for example, the lemma \myinline|sign_insert_none| states that 
for a list \myinline|φs = φ₀…φₙ| (with a slight abuse of notation) 
\myinline|(φsΛ ↩Λ φ i none).sign| is equal to \myinline|s * φsΛ.sign| where 
\myinline|s| is the sign arrived at by moving \myinline|φ| through the elements of \myinline|φ₀…φᵢ₋₁| which are contracted with some element.

\sloppy Another constructor of Wick contractions used is \myinline|join|. 
For a list of elements \myinline|φs| of \myinline|FieldOp|,
a Wick contraction \myinline|φsΛ| of \myinline|φs|, and a Wick contraction \myinline|φsucΛ| of \myinline|[φsΛ]ᵘᶜ|,
the join is defined as the Wick contraction of \myinline|φs| consisting of the contractions both in 
\myinline|φsΛ| and in \myinline|φsucΛ|. It is denoted \myinline|join φsΛ φsucΛ|.
A property of the join constructor is \myinline|join_sign_timeContract| which states that 
\myinline|(join φsΛ φsucΛ).sign • (join φsΛ φsucΛ).timeContract| is equal 
to the product of  \myinline|φsucΛ.sign • φsucΛ.timeContract| and 
\myinline|φΛ.sign • φΛ.timeContract|.

\section{Wick's theorem}

We are now in a position to give the three versions of Wick's theorem we have
formalized into Lean. Each one appears in the \myinline|<WickAlgebra>| directory of the project.
Throughout this section, let \myinline|φs| be a list of elements of \myinline|FieldOp|,
which we will consider as an element of \myinline|WickAlgebra| (by taking their product). 
 
The static version of Wick's theorem says in the algebra \myinline|WickAlgebra|
\begin{codeLong}
φs = ∑ (φsΛ : WickContraction φs.length),  
	φsΛ.sign • φsΛ.staticContract * 𝓝([φsΛ]ᵘᶜ)
\end{codeLong}
The standard version of Wick's theorem says  in the algebra \myinline|WickAlgebra|
\begin{codeLong}
𝓣(φs) = ∑ (φsΛ : WickContraction φs.length),  	
	φsΛ.sign • φsΛ.timeContract * 𝓝([φsΛ]ᵘᶜ)
\end{codeLong}
The normal-ordered version of Wick's theorem is a little more complicated. It says  in the algebra \myinline|WickAlgebra|
\begin{codeLong}
𝓣(𝓝(φs)) =  ∑ (φsΛ : {φsΛ : WickContraction φs.length // ¬ HaveEqTime φsΛ}),
	φsΛ.sign • φsΛ.timeContract * 𝓝([φsΛ]ᵘᶜ)
\end{codeLong}
The notation \myinline|∑ (φsΛ : {φsΛ : WickContraction φs.length // ¬ HaveEqTime φsΛ}), ...|
corresponds to a sum over all Wick contractions in which no two elements in the same contracted 
pair have the same time. That is, have the same time if they are position operators, or 
are both incoming or both outgoing asymptotic operators.

\section{Overview of the proof}

In what follows we illustrate the main ideas of the proof 
of the standard version of Wick's theorem and the normal-ordered version.
The proof of the static version of Wick's theorem is essentially a variation of the 
proof to the standard version. 

For more details we refer the reader to the code itself, or to 
\begin{center}
	\url{https://notes.physlean.com/QFT/Wicks-theorem/}
\end{center} 
where an interactive overview can be found. 

\subsection{Standard version of Wick's theorem}
For a list \myinline|φs = φ₀…φₙ|, 
the lemma \myinline|timeOrder_eq_maxTimeField_mul_finset|  states 
that 
\begin{codeLong}
𝓣(φs) = 𝓢(φᵢ,φ₀…φᵢ₋₁) • φᵢ * 𝓣(φ₀…φᵢ₋₁φᵢ₊₁…φₙ)
\end{codeLong}
where \myinline|φᵢ| is 
the first element of \myinline|φs| with the greatest time, and \myinline|𝓢(φᵢ,φ₀…φᵢ₋₁)| is the
sign associated with moving \myinline|φᵢ| through \myinline|φ₀…φᵢ₋₁|. 
The 
proof follows from basic properties of the supercommutator and the time-ordering map.

We then use the induction hypothesis on the list \myinline|𝓣(φ₀…φᵢ₋₁φᵢ₊₁…φₙ)| to 
rewrite it as a sum:
\begin{codeLong}
𝓣(φ₀…φᵢ₋₁φᵢ₊₁…φₙ) = ∑ (φsΛ : WickContraction (φ₀…φᵢ₋₁φᵢ₊₁…φₙ).length), 
	φsΛ.sign • φsΛ.timeContract * 𝓝([φsΛ]ᵘᶜ)
\end{codeLong}

The lemma \myinline|mul_wickTerm_eq_sum| is then used to perform the following 
	rewrite
\begin{codeLong}
φᵢ * (φsΛ.sign • φsΛ.timeContract * 𝓝([φsΛ]ᵘᶜ)) = 𝓢(φ, φ₀…φᵢ₋₁) • 
	∑ k, (φsΛ ↩Λ φ i k).sign * (φsΛ ↩Λ φ i k).timeContract * 
		𝓝([(φsΛ ↩Λ φ i k)]ᵘᶜ)
\end{codeLong}
The sum here is over all optional positions in \myinline|[φsΛ]ᵘᶜ|.
This result follows from a number of other properties and lemmas. 
Firstly, it follows from properties of the constructor \myinline|φsΛ ↩Λ φ i k| 
with respect \myinline|sign| and \myinline|timeContract|. 
An example of such a property, \myinline|sign_insert_none|, is discussed above.
Secondly, it follows from  the lemma 
\myinline|ofFieldOp_mul_normalOrder_ofFieldOpList_eq_sum|, which states that 
\begin{codeLong}
φ * 𝓝(φ₀φ₁…φₙ) = 𝓝(φφ₀φ₁…φₙ) +
	∑ i, (𝓢(φ,φ₀φ₁…φᵢ₋₁) • [anPart φ, φᵢ]ₛ) * 𝓝(φ₀…φᵢ₋₁φᵢ₊₁…φₙ).
\end{codeLong}
which itself follows from properties of the normal-ordering map and the supercommuator. 
Thirdly, it follows from a careful consideration of signs and how they interact.
Dealing with these signs takes up a considerable amount of the overall proof. 

To finish the overall proof we use the result \myinline|insertLift_sum|. This states that for a general 
function \myinline|f| from wick contractions of \myinline|φs = φ₀…φₙ| to some general 
additive commutative monoid \myinline|M|,
\begin{codeLong}
∑ (φsΛ : WickContraction (φ₀…φₙ).length), f φsΛ = 
	∑ (φsΛ : WickContraction (φ₀…φᵢ₋₁φᵢ₊₁…φₙ).length), ∑ k, f (φsΛ ↩Λ φ i k) .
\end{codeLong}
This is used to convert the two sums we now have into a sum over all Wick contractions of \myinline|φs|.

\subsection{Normal ordered version of Wick's theorem}
The proof of the normal-ordered version of Wick's theorem follows from the following 
two lemmas. 

The first lemma is the result \myinline|timeOrder_haveEqTime_split| which states that 
\begin{codeLong}
𝓣(φs) = ∑ φsΛ, φsΛ.sign • φsΛ.timeContract * (∑ φssucΛ, φssucΛ
	φssucΛ.sign • φssucΛ.timeContract * 𝓝([φssucΛ]ᵘᶜ))
\end{codeLong}
where the first term is over all Wick contractions \myinline|φsΛ| of \myinline|φs| 
which does not contract elements of \myinline|φs| with the same time, and the 
second term is over all non-empty Wick contractions \myinline|φssucΛ| of \myinline|[φsΛ]ᵘᶜ|
which only contract elements of \myinline|φsΛ| with the same time.
The proof of this result follows from an equivalence defined through
the standard version of Wick's theorem, the \myinline|join| constructor,
and the lemma \myinline|join_sign_timeContract| discussed above.

The second lemma is the results \myinline|normalOrder_timeOrder_ofFieldOpList_eq_eqTimeOnly_empty| which states 
that 
\begin{codeLong}
𝓣(𝓝(φs)) = 𝓣(φs) - ∑ φsΛ, φsΛ.sign • φsΛ.timeContract * 𝓣(𝓝([φsΛ]ᵘᶜ))
\end{codeLong}
where the sum is over all non-empty Wick contraction \myinline|φsΛ| which only 
contract equal time pairs. 
This follows from the following. 
Firstly, the static version of Wick's theorem which rewrites 
the \myinline|φs| in the first term on the left hand side.
Secondly, the property \myinline|timeOrder_staticContract_of_not_mem| which states that 
the time-ordering of a \myinline|staticContract| which has at least one contraction of a non-equal time pair
is zero. Lastly, the property \myinline|staticContract_eq_timeContract_of_eqTimeOnly| which states that
for a Wick contraction \myinline|φsΛ|  which only contracts equal-time pairs,
the static contract is equal to the time contract. 

These two combine with the induction hypothesis on the terms \myinline|𝓣(𝓝([φsΛ]ᵘᶜ))| 
to prove the result. The induction terminates since the length of the list \myinline|[φsΛ]ᵘᶜ| is less than the length of \myinline|φs|.

\section*{Acknowledgments}
I thank all those who have contributed to the PhysLean project as well as the wider Lean community.
I particularly thank Matteo Cipollina, Jeremy Lindsay, Pietro Monticone, and Zhi-Kai Pong 
for recent work on PhysLean.
I thank Tarmo Uustalu for his support throughout this project, and 
for his comments on the manuscript.
This research is supported by the project ``Icelandic advantage in computer-assisted proof''
of the Collaboration Fund of Iceland's Ministry of Higher Education, Science and Innovation.
\bibliographystyle{unsrturl}
\begin{spacing}{0.5}
\bibliography{MyBib}
\end{spacing}

\end{document}